\documentclass[11pt,fleqn]{article}
\usepackage{a4wide}
\usepackage{natbib}
\usepackage{amssymb, amsmath}
\usepackage{amsfonts}
\usepackage{amsthm}
\usepackage{graphics}
\usepackage{bm}
\usepackage{float}
\usepackage{appendix}
\usepackage{mathpazo}
\usepackage{comment}
\usepackage[multiple]{footmisc}
\usepackage{rotating}
\usepackage{subcaption}
\usepackage{booktabs}
\usepackage{threeparttable}
\usepackage{pdflscape}
\usepackage{enumerate}
\usepackage[top=2.5cm, bottom=2.5cm, left=3.0cm, right=2.5cm]{geometry}
\newtheorem{theorem}{Theorem}

\newtheorem{corollary}{Corollary}
\theoremstyle{definition}
\newtheorem{assumption}{Assumption}[section]

\newtheorem{remark}{Remark}

\numberwithin{equation}{section}
\numberwithin{lemma}{section}
\numberwithin{corollary}{section}
\numberwithin{remark}{section}
\numberwithin{theorem}{section}
\numberwithin{proposition}{section}

\begin{document}

\title{{\sc Structural Breaks in Interactive Effects Panels and the Stock Market Reaction to COVID--19}\thanks{Westerlund would like to thank the Knut and Alice Wallenberg Foundation for financial support through a Wallenberg Academy Fellowship.}}

\author{Yiannis Karavias\thanks{Corresponding author: Department of Economics, University of Birmingham, Edgbaston, B15 2TT, Birmingham, UK. E-mail address: \texttt{i.karavias@bham.ac.uk}.}\\
{\small University of Birmingham} \and Paresh Narayan\\
{\small Monash University}\and Joakim Westerlund\\
{\small Lund University}\\
{\small and}\\
{\small Deakin University}}

\maketitle

\begin{abstract}
Dealing with structural breaks is an important step in most, if not all, empirical economic research. This is particularly true in panel data comprised of many cross-sectional units, such as individuals, firms or countries, which are all affected by major events. The COVID--19 pandemic has affected most sectors of the global economy, and there is by now plenty of evidence to support this. The impact on stock markets is, however, still unclear. The fact that most markets seem to have partly recovered while the pandemic is still ongoing suggests that the relationship between stock returns and COVID--19 has been subject to structural change. It is therefore important to know if a structural break has occurred and, if it has, to infer the date of the break. In the present paper we take this last observation as a source of motivation to develop a new break detection toolbox that is applicable to different sized panels, easy to implement and robust to general forms of unobserved heterogeneity. The toolbox, which is the first of its kind, includes a test for structural change, a break date estimator, and a break date confidence interval. Application to a panel covering 61 countries from January 3 to September 25, 2020, leads to the detection of a structural break that is dated to the first week of April. The effect of COVID--19 is negative before the break and zero thereafter, implying that while markets did react, the reaction was short-lived. A possible explanation for this is the quantitative easing programs announced by central banks all over the world in the second half of March.
\end{abstract}



\section{Introduction}\label{sect:intro}

This paper considers what we believe to be a very common scenario in practice. We have in mind a researcher that seeks to infer a linear relationship between a dependent variable and a set of regressors. The data set has a panel structure, in which there are a large number of cross-sectional units, $N$, that are observed over $T$ time periods. This scenario is relevant because while the number of time periods is always limited and cannot be increased other than by the passage of time, statistical agencies keep publishing time series data for individuals, firms and countries. Thus, while $N$ is usually quite large, $T$ need not be. One of the concerns here is therefore that $T$ might not be large enough for many econometric approaches to work properly. Another concern is the presence of unobserved heterogeneity and the detrimental effect that this may have if said heterogeneity is correlated with the regressors. The main worry, however, is that the coefficients of some or indeed all of the regressors may be subject to structural change, because of some major events that may have caused the relationship to change over time. The present paper develops a toolbox that enables the researcher to test for the presence of a common structural break and, if a break is detected, to also infer the date of the break. The tools are extremely easy to implement, accommodate general forms of unobserved heterogeneity and they can be used under quite relaxed conditions on $T$, provided that $N$ is large.

Accounting for structural change has always been an important issue in economics and elsewhere. Panel data are particularly susceptible to such change, because of the large number of time series that they contain. For example, a credit crunch or debt crisis may affect the returns of many firms and households, an oil price shock may affect the output of many countries, and a fad or fashion is likely to influence the livelihood of many individuals. In our empirical application, we consider stock returns for 61 countries, which all plummeted around March 11, 2020, when COVID--19 was declared a global pandemic by the World Health Organisation (WHO). Fortunately, the panel data structure not only makes breaks likely, but it also makes for relatively easy detection. As is well known, with time series data consistent estimation of the breakpoint is not possible, but only consistent estimation of the break fraction. By contrast, in panels consistency is usually possible. Bai (2010) was among the first to make this point. He considered a highly stylized model with a breaking constant as the only regressor. The main finding is that the popular ordinary least squares (OLS) breakpoint estimator based on minimizing the sum of squared residuals is consistent as $N\to \infty$ regardless of whether $T$ is fixed or going to infinity. The accuracy of the procedure is therefore greatly enhanced when compared to the time series case.

The increased estimation accuracy is one of the advantages of using panel data. Another major advantage is the ability to deal with unobserved heterogeneity. Such heterogeneity is important in general, and it is particularly relevant in the type of noisy panels that we have in mind where typically the regressors explain only a small fraction of the variation in the dependent variable (see Capelle-Blanchard and Desroziers, 2020, in the context of COVID--19 and stock returns). This consideration motivated Horv\'{a}th and Hu\v{s}kov\'{a} (2012), Kim (2014), and Westerlund (2019) to extend the work of Bai (2010) to the case when the stochastic component of the data admits to a common factor, or ``interactive effects'', representation.\footnote{Kim (2011) also allow for common factors. This study is, however, not very constructive, in the sense that it only establishes that the factors render the OLS breakpoint estimator inconsistent.} Such representations have been shown to be very effective at capturing unobserved heterogeneity, and they are therefore very popular. It is, however, still the same breaking constant-only model that is being considered, and many models of interest involve more general regressors.

Antoch et al. (2019), Baltagi et al. (2016), Boldea et al. (2020), Hidalgo and Schafgans (2017), and Li et al. (2016) do for a general linear panel data regression model what Horv\'{a}th and Hu\v{s}kov\'{a} (2012), Kim (2014), and Westerlund (2019) do for the constant-only model. In particular, while Antoch et al. (2019), and Hidalgo and Schafgans (2017) propose tests for the presence of a structural break, Baltagi et al. (2016), Boldea et al. (2020), and Li et al. (2016) take the existence of a break as given and focus instead on the breakpoint estimation problem. But while highly complementary in terms of the methods they propose, the assumptions employed are materially different.

Antoch et al. (2019) only require $N$ to be large. However, they assume instead that the factor loadings are negligible, which means that strong forms of cross-section dependence are not permitted.\footnote{See Chudik et al. (2011) for a detailed treatment of the concepts of weak and strong cross-section dependence.} It also means that there is no need to account for the factors, and that their effect on the breakpoint estimation problem is in this sense trivial. The weak cross-sectional dependence condition is maintained also in Hidalgo and Schafgans (2017), who in addition require that $N,\,T \to\infty$ with $N/T^2 \to 0$, which in practice means that $T>> N$. Baltagi et al. (2016) do not require negligible loadings and are therefore more general in this regard. The way they do this is by applying the common correlated effects (CCE) approach of Pesaran (2006), which enables consistent estimation of (the space spanned by) the unknown factors. But then Baltagi et al. (2016) require that both $N$ and $T$ are large, which is again rarely the case in practice. Moreover, the authors only provide a consistency result and they do not consider the asymptotic distribution of the estimated breakpoint, which is necessary for the construction of confidence intervals with correct asymptotic coverage. The same critique applies to the paper of Li et al. (2016), which uses the principal components method instead of CCE to estimate factors. Boldea et al. (2020) also do not consider the asymptotic distribution of their estimated breakpoint, although in their paper $T$ is fixed. Their approach is similar to the one of Antoch et al. (2019) in the sense that the estimation is carried out while ignoring the factors. This simplicity does, however, come at a cost in terms additional restrictive assumptions. Boldea et al. (2020) do not go as far as Antoch et al. (2019) and require negligible loadings, but they do assume that the omitted variables bias caused by the factors is time-invariant, up to the breakpoint, which limits the type of factors that can be permitted.

Motivated by the above discussion, the present paper develops tools that enable researchers to both test for the presence of a common structural break, and to infer the breakpoint of an existing break. The unobserved heterogeneity is assumed to have a common factor structure that is handled by using a version of the CCE approach, which is similar yet clearly distinct from the one employed by Baltagi et al. (2016). The reason for focusing on CCE as opposed to the otherwise so popular principal components method is in part because of the extreme simplicity with which the factors are estimated in CCE, in part because CCE is valid even if $T$ is fixed (see Westerlund et al., 2019). Needless to say, this last feature, which is not exploited by Baltagi et al. (2016), is an important advantage when wanting to entertain the possibility that $T$ might not be large. The idea, which is laid out along with our model and assumptions in Section \ref{sect:mod}, is to use the cross-sectional averages of the regressors to estimate the unknown common factors, and to simply augment the regression model with these averages.

We begin by considering the problem of estimating the unknown breakpoint given that a break has occurred. This is done in Section \ref{sect:est}. Most papers in the literature are based on the OLS breakpoint estimator, and so is this paper. However, instead of minimizing the OLS residuals, which will generally lead to inconsistency because of the unattended factors (Kim, 2011), we minimize the CCE residuals. We focus on the results for the case when the magnitude of the break is bounded from above and below, although we also allow diverging and shrinking breaks. Moreover, $T$ can be fixed or tending to infinity. According to the results, the proposed breakpoint estimator is consistent as $N\to\infty$ with $T$ fixed or as $N,\,T\to \infty$ with $T/N\to 0$, and the rate of convergence is given by $1/N$. The asymptotic distribution of the breakpoint estimator is obtained under the same set of conditions on $N$ and $T$, and is used to construct confidence intervals for the true breakpoint. As far as we are aware, this paper is the first to provide the rate of convergence and asymptotic distribution in the presence of common factors, and it is the first to establish consistency when $T$ is fixed.

While in Section \ref{sect:est} we assume that a break has occurred, in Section \ref{sect:tests} we instead consider the problem of testing for the presence of a break. While very common in the literature, CUSUM-based test statistics like the one of Hidalgo and Schafgans (2017) can suffer from low power in certain directions (see, for example, Andrews, 1993). In this paper, we therefore follow the recommendation of Andrews (1993), and consider two Wald-type test statistics. One is designed to test the null hypothesis of no break against the alternative of a known breakpoint, while in the other the same null is tested against the alternative of a break at some unknown date. To the best of our knowledge, these tests are the first that enable break testing in the presence of common factors. The asymptotic analysis reveals that, while the consistency and asymptotic distribution of the breakpoint estimator only require $N\to\infty$, unless $N,\,T\to \infty$ with $T/N\to 0$, the asymptotic distribution of the Wald test statistics are generally not free of nuisance parameters. Hence, in terms of the size of $T$, testing for the presence of a structural break is more demanding than estimating the breakpoint. This is what theory tells us. According to the Monte Carlo results reported in the online appendix, however, the new toolbox tends to perform well even if $T$ is as small as 10, provided that $N$ is large enough. Hence, even if in theory the Wald tests require $T$ to be large, in small samples this requirement does not seem very critical.

Section \ref{sect:application} is concerned with our empirical application to the relationship between stock returns and COVID--19, which is motivated in part by the many recent calls for econometric research into the effects of the pandemic (see, for example, the recent special issue of Journal of Econometrics), in part by existing empirical research. By the end of February, 2020, COVID--19 had led to a world-wide drop in demand, which in turn brought down investment and employment. While stock markets initially reacted to news of the pandemic by losing substantial value, they quickly regained the vast majority of this loss. The fact that this rebound took place even though the number of new cases and deaths were still rising is suggestive of structural change. Most studies of the stock market reaction to COVID--19 either ignore this possibility altogether or split their sample into subperiods based on major events (see, for example, Capelle-Blanchard and Desroziers, 2020, and Ramelli and Wagner, 2020). This means that the breaks are treated as known, if treated at all, which is risky, as misplaced breaks are just as problematic as omitted breaks. In the empirical application of the present paper, we offer a more general treatment. This is done by applying the new toolbox to a sample covering 61 countries across 38 weeks, from January 3 to September 25, 2020, which means that $T$ is relatively small. According to the results, the COVID--19-stock return relationship has been affected by the presence a structural break in the first week of April, at about the same time as most central banks announced that they were going to intervene to save the global economy from collapse. While before the break stock markets reacted significantly to news about the pandemic, after the break stock markets became insensitive to such news. This suggests that central banks play a central role in shaping stock market behaviour in pandemics. 

Section \ref{sect:concl} concludes the paper. All proofs are provided in the online appendix.

\section{Model}\label{sect:mod}

In this paper, we consider the following linear panel data model with a structural break at time $b$:
\begin{equation}
y_{i,t}= \beta'x_{i,t}+\delta'z_{i,t}(b)+e_{i,t},
\label{eq:y}
\end{equation}
where $i=1,...,N$ and $t=1,...,T$ index the cross-sectional units and time periods, respectively. The $k \times 1$ vector $x_{i,t}$ contains the regressors and the $r \times 1$ vector $z_{i,t}$ is defined as
\begin{equation}
z_{i,t}(b)=R'x_{i,t} \mathbb{I}(t>b),
\label{eq:z}
\end{equation}
where $\mathbb{I}(t>b)$ is the indicator function taking the value one when $t>b$ and zero otherwise, and $R$ is an $k \times r$ selection matrix of zeros and ones with full column rank $r$ that picks out the elements of $x_{i,t}$ whose coefficients are subject to structural change. For example, if $k > r$ and $R=(0_{r \times (k - r)}',I_{r})'$, then \eqref{eq:y} is a partial structural change model in which only the $r$ last regressors in $x_{i,t}$ appear in $z_{i,t}(b)$. If, on the other hand, $k = r$, then $R= I_{r}$, and so the model is one of pure structural change. In the empirical application of Section \ref{sect:application}, $y_{i,t}$ is stock returns for country $i$ in week $t$, and $x_{i,t}$ is comprised of controls and COVID--19 related variables, where the coefficients of the COVID--19 variables may be breaking. As we explain in that section, the model can easily be generalized to include multiple structural changes. The coefficient vectors $\beta$ and $\delta$ are of dimension $k \times 1$ and $r \times 1$, respectively.

The error $e_{i,t}$ is assumed to admit to a factor structure, which means that it is allowed to be correlated across $i$. Specifically,
\begin{equation}
e_{i,t}=\gamma_{i}'f_{t}+\varepsilon_{i,t},
\label{eq:e}
\end{equation}
where $f_{t}$ and $\gamma_{i}$ are $m \times 1$ vectors of common factors and factor loadings, respectively, and $\varepsilon_{i,t}$ is an idiosyncratic error term. In our empirical application, the presence of $f_t$ in \eqref{eq:e} is just natural because many well-known models in finance, like the capital asset pricing (CAPM) and Fama--French (FF) three factor models, imply that returns should have a linear factor structure. In this section and the next, we assume that all the factors are unknown. In Section \ref{sect:application}, we demonstrate how the toolbox can be implemented when some of the factors are observed, as when CAPM holds and one has data on (world) market returns.

We want to entertain the possibility that the factors are correlated with the regressors. We therefore follow Pesaran (2006) and assume that
\begin{equation}
x_{i,t}= \Gamma_{i}' f_{t} +v_{i,t},
\label{eq:x}
\end{equation}
where $\Gamma_{i}$ is a $m \times k$ factor loading matrix and $v_{i,t}$ is a $k \times 1$ vector of idiosyncratic errors.

For later use, it is convenient to write the above model in matrix form by stacking the time series observations for each cross-section. The stacked version of \eqref{eq:y} is given by
\begin{equation}
y_i= X_i\beta+Z_i(b)\delta + e_{i},
\end{equation}
where $y_i=(y_{i,1},...,y_{i,T})'$ and $e_{i}=(e_{i,1},...,e_{i,T})'$ are $T\times 1$, $X_i=(x_{i,1}',...,x_{i,T}')'$ is $T\times k$, and $Z_i(b)=(z_{i,1}(b)',...,z_{i,T}(b)')'$ is $T\times r$. Note that because $z_{i,1}(b)=...=z_{i,b}(b) = 0_{r \times 1}$, $Z_i(b)$ can be written as $Z_i(b)=(0_{r \times 1}',...,0_{r \times 1}',x_{i,b+1}'R,...,x_{i,T}'R)'= X_i(b) R$, where $X_i(b) = (0_{k \times 1}',...,0_{k \times 1}',x_{i,b+1}',...,x_{i,T}')'$. Also,
\begin{equation}
e_{i} = F \gamma_{i} +\varepsilon_{i},
\end{equation}
where $F=(f_{1}',...,f_{T}')'$ and $\varepsilon_{i}=(\varepsilon_{i,1},...,\varepsilon_{i,T})'$ are $T\times m$ and $T\times 1$, respectively. The stacked version of \eqref{eq:x} is given by
\begin{equation}
X_i= F \Gamma_{i}+ V_{i},
\end{equation}
where $V_{i}=(v_{i,1}',...,v_{i,T}')'$ is $T\times k$.

The model assumptions depend to a large extent on whether we are estimating the breakpoint or if we are testing for its existence. Assumptions \ref{ass:err} and \ref{ass:f} will, however, be maintained throughout this paper.

\begin{assumption}\label{ass:err}\
\begin{enumerate}[(a)]
\item $v_{i,t}$ is a covariance stationary process that is independent across $i$ with absolutely summable autocovariances, $E(v_{i,t}) = 0_{k\times 1}$, $E(v_{i,t}v_{i,t}')=\Sigma_{v,i}$ and $E(\|v_{i,t}\|^4)< \infty$, where $\|A\| = \sqrt{\mathrm{tr}(A'A)}$ is the Frobenius norm of any matrix $A$.

\item $\varepsilon_{i,t}$ is a covariance stationary process that is independent across $i$ with absolutely summable autocovariances, $E(\varepsilon_{i,t})=0$, $E( \varepsilon_{i,t}^2 ) =\sigma_{\varepsilon,i}^{2} > 0$ and $E(\varepsilon_{i,t}^4) < \infty$.

\item $\varepsilon_{i,t}$ and $v_{j,s}$ are independent for all $i$, $j$, $s$ and $t$.
\end{enumerate}
\end{assumption}

\begin{assumption}\label{ass:f}\
\begin{enumerate}[(a)]
\item $T^{-1}F'F$ is positive definite with probability approaching one (w.p.a.1) for all $T$.

\item $E(\|f_t\|^4) < \infty$.

\item $f_t$ is independent of $\varepsilon_{i,s}$ and $v_{i,s}$ for all $i$, $s$ and $t$.
\end{enumerate}
\end{assumption}

Assumption \ref{ass:err} is standard in the interactive effects literature (see, for example, Baltagi et al., 2016). The only exception known to us is Baltagi et al. (2017). They do not allow for cross-section dependence, but they do allow $x_{i,t}$ and $\varepsilon_{i,t}$ to be unit root non-stationary. We allow for serial correlation and possibly even unit roots in $f_{t}$ (more later) and hence in $x_{i,t}$ (and $y_{i,t}$), but not in $v_{i,t}$ and $\varepsilon_{i,t}$. Bai (1997a) has shown that the existence of both serially correlated errors and lagged dependent variables leads to inconsistent estimation of the break date. Assumption \ref{ass:err} (c) therefore assumes that $x_{i,t}$ is strictly exogenous. Without unit roots Assumption \ref{ass:err} (a)--(c) are the same as Assumptions 1--3 in Baltagi et al. (2017). Assumption \ref{ass:f} (a) and (b) are met if $f_t$ is stationary and not collinear, which is again a standard requirement in the literature (see Baltagi et al., 2016, Assumption 8). Stationarity is not necessary, though. Note in particular how stationarity is not required if $T$ is fixed. In fact, $f_t$ does not even have to be stochastic but can also be deterministic. Assumption \ref{ass:f} (c) is an identifying condition that is not particularly restrictive. It ensures that $f_t$ is the only source of cross-section dependence.

\section{Breakpoint estimation}\label{sect:est}

Let us denote by $b_0$ the true value of $b$. The purpose of this section is to make inference regarding this parameter.

\begin{assumption}\label{ass:break}
$b_0 \in \mathcal{B}=[r,T-r-1]$.
\end{assumption}

\begin{assumption}\label{ass:load}\
\begin{enumerate}[(a)]
\item $\mathrm{rank}(\bar{\Gamma}) = m\leq k$ for all $N$, including $N\to\infty$.

\item $\|\gamma_{i}\| < \infty$ and $\|\bar{\Gamma}\| < \infty$.
\end{enumerate}
\end{assumption}

Assumption \ref{ass:break} requires only that each regime contains at least as many observations as the number of free parameters. It is therefore very general. Baltagi et al. (2016) and Westerlund (2019) allow for common factors in very much the same way as we do. However, they require that the loadings follow certain probability laws, and that they are independent of all other random elements of the model. In this section, we treat the loadings as fixed, which means that we do not make any assumption regarding their distribution or their correlation with the other random elements of the model. The main restriction is that $x_{i,t}$ must load on the same factors as $y_{i,t}$, and that the number of regressors must be at least as large as the number of factors. This ensures that the factors can be estimated by applying CCE to $x_{i,t}$, as we will now explain.

Unlike in Antoch et al. (2019), where the factor loadings are assumed to be negligible, under our conditions valid inference on $b_0$ is not possible without proper accounting for $f_t$. The reason is that the factors make $x_{i,t}$ correlated with $e_{i,t}$, which means that \eqref{eq:y} cannot be estimated consistently using OLS. However, we note that $x_{i,t}$ has a pure factor model representation, suggesting that the factors can be estimated using methods designed for such models. In this paper, we follow Baltagi et al. (2016), and use the CCE approach of Pesaran (2006), which is based on using the cross-sectional average of the observables to estimate the space spanned by $f_t$. The difference is that we do not include the cross-sectional average of $y_{i,t}$, which in the current context is uninformative regarding $f_t$. This is shown in the online appendix. Hence, in contrast to Baltagi et al. (2016), in the current paper we only use $\bar{x}_t$, where $\bar{A}_{t}=N^{-1}\sum_{i=1}^{N} A_{i,t}$ is the cross-sectional average of any variable $A_{i,t}$. In view of \eqref{eq:x}, this average can be written as
\begin{equation}\label{eq:xbar}
\bar{x}_{t} =\bar{\Gamma}'f_{t}+\bar{v}_{t}.
\end{equation}
Let $A^+$ denote the Moore--Penrose inverse of any matrix $A$. If Assumption \ref{ass:load} is true, so that $\bar{\Gamma}$ has full row rank, the Moore--Penrose inverse of $\bar{\Gamma}$ is given by $\bar{\Gamma}^+ = \bar{\Gamma}'( \bar{\Gamma}\bar{\Gamma}' )^{-1}$. Hence, $\bar{\Gamma}\bar{\Gamma}^+ = I_m$, which in turn means that \eqref{eq:xbar} can be solved for $f_t$ by left-multiplication by $\bar{\Gamma}^{\prime+}$. It follows that if Assumption \ref{ass:err} is also true, so that $\|\bar{v}_{t}\| = o_p(1)$, then
\begin{equation} \label{eq:fsol}
\|\bar{\Gamma}^{\prime+}\bar{x}_{t} - f_{t}\| = \|\bar{\Gamma}^{\prime+}\bar{v}_{t}\| \le \|\bar{\Gamma}^{\prime+}\|\|\bar{v}_{t}\| = o_p(1).
\end{equation}
We say that $\bar{x}_{t}$ is ``rotationally consistent'' for $f_{t}$, because it is consistent up to an invertible rotation matrix. Hence, by augmenting \eqref{eq:y} with $\bar{x}_{t}$, provided that $N$ is large, we can control for $f_t$, and in this way break the correlation between the regressors and the error term.

Define $M_{\bar{X}}=I_{T}-\bar{X}(\bar{X}'\bar{X})^{-1}\bar{X}'$, where $\bar{X}=N^{-1}\sum_{i=1}^{N} X_{i} = (\bar{x}_{1},...,\bar{x}_{T})'$, and let $\tilde{A}_i = M_{\bar{X}} A_i$ for any $T$-rowed matrix $A_i$. The augmented model to be estimated can now be written as
\begin{equation}\label{eq:ytstack}
\tilde{y}_i =\tilde{X}_i \beta + \tilde{Z}_{i}(b_0)\delta + \tilde{e}_{i}.
\end{equation}
This model can be stacked also over the cross-section, giving
\begin{equation}\label{eq:ystack}
\tilde{Y}= \tilde{X}\beta+\tilde{Z}(b_0)\delta +\tilde{E},
\end{equation}
where $\tilde{Y}=(\tilde{y}_{1}',...,\tilde{y}_{N}')'$ and $\tilde{E}=(\tilde{e}_{1}',...,\tilde{e}_{N}')'$ are $NT\times 1$, $\tilde{X}=(\tilde{X}_{1}',...,\tilde{X}_{N}')'$ is $NT\times k$ and $\tilde{Z}(b_0)=(\tilde{Z}_{1}(b_0)',...,\tilde{Z}_{N}(b_0)')'$ is $NT\times r$. Let us further introduce $M_{\tilde{X}}=I_{NT}-\tilde{X}(\tilde{X}'\tilde{X})^{-1}\tilde{X}'$. With $b_0$ known, the CCE estimator of $\delta$, which is identically the OLS estimator obtained from \eqref{eq:ystack}, and the associated sum of squared residuals are given by
\begin{align}
\hat{\delta}(b_0) &= (\tilde{Z}(b_0)'M_{\tilde{X}}\tilde{Z}(b_0))^{-1} \tilde{Z}(b_0)'M_{\tilde{X}}\tilde{Y},\\
SSR(b_0) &= (\tilde{Y}- \tilde{Z}(b_0)\hat{\delta}(b_0))'M_{\tilde{X}} (\tilde{Y}-\tilde{Z}(b_0)\hat{\delta}(b_0)).
\end{align}
Of course, in many scenarios of empirical relevance, $b_0$ is not known. The estimator that we will use in its stead is obtained by minimizing $SSR(b)$ over all possible values of $b$;
\begin{equation}
\hat b = \arg\min_{ b \in \mathcal{B}} SSR(b).
\end{equation}

We begin by showing that $\hat b$ is consistent. For this to be possible, however, in addition to Assumptions \ref{ass:err}--\ref{ass:load}, we need to ensure that the inverse appearing in $\hat{\delta}(b)$ is well-behaved. This is where Assumption \ref{ass:regr} comes in. It demands that the regressors in $x_{i,t}$ have enough variation across both $i$ and $t$ after projecting out all variation that can be explained by $f_t$. This rules out cross-section-invariant regressors in $x_{i,t}$.

\begin{assumption}\label{ass:regr}\
\begin{enumerate}[(a)]
\item $(NT)^{-1}\tilde{X}'\tilde{X}$ is positive definite w.p.a.1 for all $N$ and $T$.

\item $(NT)^{-1}\tilde{Z}(b)'M_{\tilde{X}}\tilde{Z}(b)$ is positive definite w.p.a.1 for all $b\in \mathcal{B}$, $N$ and $T$.
\end{enumerate}
\end{assumption}

We are now ready to state our first main result.

\begin{theorem}\label{thm:rates}
Suppose that Assumptions \ref{ass:err}, \ref{ass:f} and \ref{ass:break}--\ref{ass:regr} are met. Then, the following results hold:
\begin{enumerate}[(a)]
  \item If $N\to\infty$ and $T$ is fixed, or as $N,\,T\to\infty$ with $\sqrt{N}\| \delta\|\to\infty$ when $m = k$,
\begin{align}
|b_0- \hat b| = O_p(\| \delta\|^{-2} N^{-1}).
\end{align}

  \item If $N,\,T\to\infty$ with $\sqrt{N}\| \delta\|\to\infty$ and $T/(N\| \delta\|^2) \to 0$ when $m < k$,
\begin{align}
|b_0-\hat b| = o_p(\| \delta\|^{-2}N^{-1}).
\end{align}
\end{enumerate}
\end{theorem}

Theorem \ref{thm:rates} states that $\hat{b}$ is consistent and that the rate of convergence is $\| \delta\|^{-2} N^{-1}$ or better. The fact that consistency is possible even if $T$ is fixed is very useful in practice, because it means that breaks can be detected very quickly. When $T\to\infty$ it matters whether $m < k$ or $m=k$. Note in particular how the rate of convergence is faster when $m < k$ then when $m = k$, and that this is true even if $T/(N\| \delta\|^2) \to 0$ under $m = k$, so that the conditions for (a) and (b) are the same. The reason is that when $m < k$, unlike what one would expect based on standard theory for regressions in stationary variables, the effect of the redundant cross-section averages contained in $\bar{X}$ are not negligible but impact the asymptotic theory in very much the same way as unit root regressors do in a spurious regression. Moreover, the redundant averages are correlated with the breaking regressors in $Z_{i}(b)$, and this increases the signal coming from $\tilde{Z}_{i}(b)$. As far as we are aware, this is the first time redundant regressors have been shown to lead to increased accuracy in breakpoint estimation.

If we are not interested in the distinction between $m < k$ or $m=k$, the results contained in Theorem \ref{thm:rates} can be stated as in Corollary \ref{cor:rates}.

\begin{corollary}\label{cor:rates}
Suppose that conditions of Theorem \ref{thm:rates} are met, and that $T/(N\| \delta\|^2) \to 0$ and $\sqrt{N}\| \delta\|\to\infty$. Then, as $N\to \infty$ with $T$ fixed, or as $N,\,T\to\infty$,
\begin{align}
|b_0- \hat b| = O_p(\| \delta\|^{-2} N^{-1}).
\end{align}
\end{corollary}

\begin{remark}\label{rem:baltagietal}
As already mentioned, Baltagi et al. (2016) consider a model that is very similar to ours and that is estimated using CCE. They show that $\hat b$ is consistent for $b_0$; however, they only consider the case when $N,\,T\to \infty$, and they do not provide the rate of convergence. Moreover, the proof that they provide is based on the assumption that $(\bar y_{t},\bar{x}_{t}')'$ is rotationally consistent for $(f_t',f_{t}'\mathbb{I}(t>b))'$, which we show in the online appendix not to be correct. Bai (2010) is the only other paper that we are aware of that proves consistency under both fixed and large $T$; however, his model is very simple in the sense that it does not contain any regressors except for a breaking mean. Under stationarity, the model considered by Baltagi et al. (2017) is very similar to our but without interactive effects. The rate given in Corollary \ref{cor:rates} is consistent with the one given in their Theorem 2.
\end{remark}

\begin{remark}\label{rem:smallbreaks}
Corollary \ref{cor:rates} requires that $T/(N\| \delta\|^2) \to 0$ and $\sqrt{N}\| \delta\|\to\infty$. The latter condition is similar to Assumption 2 in Bai (2010), and is tantamount to requiring $\delta = N^{-\alpha}\delta_0$ with $\alpha < 1/2$ and $\|\delta_0\| \in (0,\infty)$. Hence, while we allow for it, we do not require $\|\delta \| \to 0$, which is in contrast to studies such as Antoch et al. (2019), where the magnitude of the break must be shrinking. The condition that $T/(N\| \delta\|^2) \to 0$, which is similar in spirit to Assumption 2 in Baltagi et al. (2016), restricts the relative rate of expansion of $N$ and $T$, and is only needed when $T$ is large. For example, if $\| \delta\| = O(1)$, then we require that $T/N \to 0$, as otherwise the error coming from the estimation of the factors will tend to accumulate as we sum over time. We also see that the larger is $\| \delta\|$, the weaker the condition on $T/N$, as to be expected, because a larger break is easier to discern. If $T$ is fixed, then $T/(N\| \delta\|^2) \to 0$ is implied by $\sqrt{N}\| \delta\|\to\infty$.
\end{remark}

As Corollary \ref{cor:rates} makes clear, provided that $T/(N\| \delta\|^2) \to 0$ and $\sqrt{N}\| \delta\|\to\infty$, consistency holds irrespectively of whether $m = k$ or $m < k$, which is of course very useful in practice, as $m$ is unknown here. This invariance is reflected also in the asymptotic distribution of the estimated break date, as our next theorem, Theorem \ref{thm:dist}, makes clear. Before we take the theorem, however, we need to introduce a few more conditions, which are given in Assumption \ref{ass:mom}.

\begin{assumption}\label{ass:mom}\
\begin{enumerate}[(a)]
\item $E(\varepsilon_{i,t}\varepsilon_{i,s}) = 0$ for all $i$ and $t\ne s$.

\item $N^{-1}\sum_{i=1}^N \sigma_{\varepsilon,i}^{2} E(x_{i,t}x_{i,t}') \to \Phi_X$ as $N\to\infty$ for all $t$.

\item $N^{-1} \sum_{i=1}^N E(x_{i,t}x_{i,t}') \to \Omega_X$ as $N\to\infty$ for all $t$, where $\Omega_X$ is positive definite.
\end{enumerate}
\end{assumption}

Assumption \ref{ass:mom} is restrictive, but is similar to the conditions used in the previous literature (see, for example, Bai, 1997a, 2010). It demands that $\varepsilon_{i,t}$ is serially uncorrelated and that the large-$N$ moments of $x_{i,t}$ do not depend on time. While indeed quite strong, because of the presence of $f_t$, the first condition does not rule out serial correlation in $e_{i,t}$. The second requirement is stronger than necessary, and can be relaxed to accommodate moments that are constant within break regimes but potentially varying between regimes, as in, for example, Bai (1997a), and Perron and Yamamoto (2013).

\begin{theorem}\label{thm:dist}
Suppose that Assumptions \ref{ass:err}, \ref{ass:f} and \ref{ass:break}--\ref{ass:mom} are met, and that $T/(N\| \delta\|^2) \to 0$ and $\sqrt{N}\| \delta\|\to\infty$. Then, as $N\to \infty$ with $T$ fixed, or as $N,\,T\to\infty$ with $T/N\to 0$,
\begin{align}
\frac{(\delta'R'\Omega_XR\delta)^2}{\delta'R'\Phi_XR\delta} N (\hat b - b_0) \to_d \arg\max_{v\in [-V, V] \subset \mathbb{R}}\left(-\frac{|v|}{2}  + B(v) \right)
\end{align}
where $B(v)$ is standard two-sided Brownian motion on $v\in [-V, V] \subset \mathbb{R}$.\footnote{The two-sided Brownian motion $B(v)$ satisfies $B(0) = 0$, and $B(v)= B_1(v)$ for $v > 0$ and $B(v)= B_2(v)$ for $v < 0$, where $B_1(v)$ and $B_2(v)$ are two independent standard Brownian motions.}
\end{theorem}

\begin{remark}\label{rem:confint}
Most papers stop at consistency and do not report the asymptotic distribution of the estimated breakpoint. There are, however, a few exceptions. The study of Bai (2010) is one of them. He provides the asymptotic distribution of the estimated breakpoint for a model with a break-in-mean only, and no other regressors or error cross-section dependence. Kim (2011) extends this analysis to a model that allows for a break in both the mean and trend, and where the errors have a factor structure. Baltagi et al. (2018) allow for more general regressors. However, they assume instead that the errors are cross-sectionally independent. All three papers require that $T$ is large in their distributional analyses. As far as we are aware, the asymptotic distribution reported in Theorem \ref{thm:dist} is the first to allow for general regressors and common factors in panels where only $N$ is required to be large.
\end{remark}

Theorem \ref{thm:dist} can be used to construct confidence intervals for $b_0$ with asymptotically correct coverage. Under Assumption \ref{ass:mom}, consistent estimators of $\Omega_X$ and $\Phi_X$ can be constructed in the following obvious manner:
\begin{align}
\hat \Omega_X &= \frac{1}{NT}\sum_{i=1}^N X_{i}'X_{i}, \\
\hat \Phi_X &= \frac{1}{NT}\sum_{i=1}^N \hat\sigma_{\varepsilon,i}^2 X_{i}' X_{i},
\end{align}
where $\hat\sigma_{\varepsilon,i}^2 = T^{-1}\hat \varepsilon_i'\hat \varepsilon_i$ with the $T\times 1$ vector $\hat \varepsilon_i$ being the $i$-th block of the $NT \times 1$ vector $\hat \varepsilon = (\hat \varepsilon_1',...,\hat \varepsilon_N')' = M_{\tilde{X}} (\tilde{Y}-\tilde{Z}(\hat b)\hat{\delta}(\hat b))$. The probability density function of $\arg\max_ v (-|v|/2  + B(v))$ is known analytically and is given in Bai (1997a). Let us denote by $c_\alpha$ the $(1-\alpha/2)$-th percentile of this distribution function, and let $\lfloor x\rfloor$ be the integer part of $x$. In analogy to Bai (1997a), an asymptotically correctly sized $100(1-\alpha)\%$ confidence interval for $b_0$ can now be constructed as
\begin{align}
\left[ \hat b - \left\lfloor c_\alpha \cdot \frac{\hat \delta(\hat b)'R'\hat \Phi_X R \hat \delta(\hat b)}{N(\hat \delta(\hat b)'R'\hat \Omega_X R \hat \delta(\hat b))^2} \right\rfloor -1,\, \hat b + \left\lfloor c_\alpha \cdot \frac{\hat \delta(\hat b)'R'\hat \Phi_X R \hat \delta(\hat b)}{N(\hat \delta(\hat b)'R'\hat \Omega_X R \hat \delta(\hat b))^2} \right\rfloor + 1\right].
\end{align}

\section{Break testing}\label{sect:tests}

Testing for the existence of a structural break is a key first step before estimating the date of the break. In terms of the parameters of \eqref{eq:y}, the null hypothesis of no structural change is given by $H_{0}:\delta=0_{r\times 1}$. The alternative hypothesis can be formulated in (at least) two ways. We begin by considering the alternative that there is a single structural change ($\delta\neq 0_{r\times 1}$) at a given date $b$, which may or may not be equal to $b_0$. This hypothesis, henceforth denoted $H_{1}(b)$, can be tested using the following Wald test statistic:
\begin{equation}
W(b)= NT \hat \delta(b)' \hat \Sigma_\delta(b)^{-1} \hat \delta(b),\label{eq:w}
\end{equation}
where $\hat \Sigma_\delta(b)$ is a consistent estimator of the asymptotic covariance matrix of $\hat \delta(b)$, whose construction will be discussed later. Interestingly, $W(b)$ will not have the expected asymptotic chi-squared distribution with $r$ degrees of freedom, henceforth denoted $\chi^2(r)$, under $H_0$. The intuition behind this result goes as follows. As already pointed out, because of the presence of $f_t$ in both \eqref{eq:e} and \eqref{eq:x}, $\tilde X_i$ is generally endogenous. The exception is in large-$N$ samples, since here $\bar X$ is rotationally consistent for $F$, and in this sense $\tilde X_i$ is ``asymptotically exogenous''. The problem is that while the use of $M_{\bar X}$ takes care of the factors in $X_i$, it does not take care of those in $Z_i(b)$, which are breaking. This is a problem because it means that while $\tilde X_i$ is asymptotically exogenous, $\tilde Z_i(b)$ is not, which in turn invalidates inference based on $W(b)$. Because of the consistency of $\hat b$, in Section \ref{sect:est} the endogeneity of $\tilde Z_i(b)$ was not an issue. Of course, if we knew that there was a break present, as we did in Section \ref{sect:est}, there would be no need to test for it in the first place. The situation considered here is therefore quite different and this requires some changes.

The first change we make when compared to Section \ref{sect:est}, which is quite natural given the discussion of the last paragraph, is to replace $\bar X$ with $\bar H(b)=(\bar X,\bar{Z}(b))$ and $\tilde{A}_i = M_{\bar{X}} A_i$ with $\tilde{A}_i(b) = M_{\bar{H}(b)} A_i$. The definitions of $\hat \delta(b)$ and $W(b)$ are adapted accordingly. The idea here is that by augmenting $\bar X$ with $\bar{Z}(b)$, we can eliminate the factors in both $X_i$ and $Z_i(b)$, which means that the endogeneity issue is gone. For this to happen, however, we need some additional assumptions. In order to appreciate this, note that
\begin{align}
\left(\begin{array}{c}
           x_{i,t} \\
           z_{i,t}(b) \end{array}\right) & = \left(\begin{array}{cc}
           \Gamma_{i}' & 0_{k\times m} \\
           0_{r\times m} & R'\Gamma_{i}'
                                     \end{array}\right)\left(\begin{array}{c}
           f_t \\
           f_{t}\mathbb{I}(t>b) \end{array}\right)+ \left(\begin{array}{c}
           v_{i,t}\\
           R'v_{i,t}\mathbb{I}(t>b) \end{array}\right),
\end{align}
where $(f_t',f_{t}'\mathbb{I}(t>b))'$ are the factors in $(\bar x_{i,t}',\bar{z}_{i,t}(b)')'$. Hence, provided that $\mathrm{rank}(\bar{\Gamma}) = \mathrm{rank}(\bar{\Gamma}R) = m$, such that the $(k+r)\times 2m$ matrix
\begin{align}
\left(\begin{array}{cc}
           \bar{\Gamma}' & 0_{k\times m} \\
           0_{r\times m} & R'\bar{\Gamma}'
\end{array}\right)
\end{align}
has full column rank $2m \leq k+r$, analogous to the discussion of Section \ref{sect:mod}, $(\bar x_{t}',\bar{z}_{t}(b)')'$ is rotationally consistent for $(f_t',f_{t}'\mathbb{I}(t>b))'$. We also need to restrict the type of heterogeneity that can be permitted in $\gamma_i$. The way we do this is by assuming that $\gamma_i$ admits to a random coefficient representation, similarly to, for example, Pesaran (2006) and Karabiyik et al. (2017). Assumption \ref{ass:load2} below replaces Assumption \ref{ass:load} and is enough to ensure that the effect of the estimation of $f_t$ is asymptotically eliminated.

\begin{assumption}\label{ass:load2}\
\begin{enumerate}[(a)]
\item $\mathrm{rank}(\bar{\Gamma}) = m\leq k$ and $\mathrm{rank}(\bar{\Gamma}R)=m \leq r$ for all $N$, including $N\to\infty$.

\item $\|\bar{\Gamma}\| < \infty$.

\item $\gamma_i$ is independent across $i$, and of $\varepsilon_{j,t}$, $v_{j,t}$ and $f_t$ for all $i$ and $j$ with $E(\gamma_i) = \gamma$ and $E(\|\gamma_i\|^2) < \infty$.
\end{enumerate}
\end{assumption}

Another difference when compared to Section \ref{sect:est}, where $T$ could be allowed to be fixed, is that here both $N$ and $T$ have to be large. The basic reason for this is that while before we took the break as given, which meant that we could make use of the consistency of $\hat b$ to construct asymptotically valid confidence intervals, here we do not know if there is a break present and so we cannot rely on said consistency. This means that the asymptotic distribution of $W(b)$ is generally not nuisance parameter free. The main exception is if $T$ is large.\footnote{It is possible to conduct asymptotically valid inference based on $W(b)$ even if $T$ is fixed. However, this requires placing very strong assumptions regarding the time-variation of the factors, which is something that we would like to avoid.} But if $T$ is large, we can relax the serial uncorrelatedness and time invariant moment conditions of Assumption \ref{ass:mom2}. We also require that Assumption \ref{ass:regr} holds when $\bar H(b)$ is used in place of $\bar X$.

\begin{assumption}\label{ass:mom2}\
\begin{enumerate}[(a)]
\item $(NT)^{-1}\sum_{i=1}^N E(V_i'\Sigma_{\varepsilon,i}V_i) \to \Phi_V$ as $N,\,T\to\infty$, where $\Sigma_{\varepsilon,i} = E(\varepsilon_i\varepsilon_i')$.

\item $(NT)^{-1} \sum_{i=1}^N E(V_i'V_i) \to \Omega_V$ as $N,\,T\to\infty$, where $\Omega_V$ is positive definite.
\end{enumerate}
\end{assumption}

\begin{assumption} \label{ass:regr2}\
\begin{enumerate}[(a)]
\item $(NT)^{-1}\tilde{X}(b)'\tilde{X}(b)$ is positive definite w.p.a.1 for all $N$ and $T$.

\item $(NT)^{-1}\tilde{Z}(b)'M_{\tilde{X}(b)}\tilde{Z}(b)$ is positive definite w.p.a.1 for all $b\in \mathcal{B}'$, $N$ and $T$.
\end{enumerate}
\end{assumption}

A third and final difference when compared to Section \ref{sect:est} is about the range of values considered for $b$. In Section \ref{sect:est}, we only required that $b \in \mathcal{B}$, which meant that in the large-$T$ case $b/T$ could take on any value in $[0,1]$. Here this is not possible, for it is only when $b/T$ is bounded away from zero and one that $W(b)$ converges in distribution (see Andrews, 1993, for a discussion). In this section, we therefore assume that $b\in \mathcal{B}'$, where
\begin{equation}
\mathcal{B}' = \{b: b = \lfloor \tau T \rfloor ~ \text{with} ~ \tau\in \mathcal{T} = [\epsilon,1-\epsilon] ~ \text{and} ~  \epsilon > 0 \}.
\end{equation}
The main implication of this in practice is that we have to truncate, or ``trim'', the range of values considered for $b$ at both beginning and end. A very common way to do this is to set $\epsilon = 0.15$, so that the first and last 15\% of the observations are discarded (see, for example, Andrews, 1993, and Bai, 1997a). The condition that $b/T$ should bounded away from zero and one should hold for all $b$, including $b_0$. The following assumption reflects this.

\begin{assumption}\label{ass:break2}
$b_0 = \lfloor \tau_0T \rfloor$, where $\tau_0 \in \mathcal{T}_0 \subset \mathcal{T}$.
\end{assumption}

We now have all the conditions we need in order to obtain the asymptotic distribution of $W(b)$.

\begin{theorem}\label{thm:dist2}
Suppose that $H_0$ holds, and that Assumptions \ref{ass:err}, \ref{ass:f} and \ref{ass:load2}--\ref{ass:break2} are met. Then, uniformly in $b\in \mathcal{B}'$, as $N,\,T\to \infty$ with $T/N\to 0$,
\begin{equation}
W(b) \to_d \frac{[J(\tau)-\tau J(1)]'[J(\tau)-\tau J(1)]}{\tau(1-\tau)},
\end{equation}
where $J(\tau)$ is a $r\times 1$ vector standard Brownian motion on $\tau\in \mathcal{T} \subset (0,1)$.
\end{theorem}

Because $J(\tau)$ is a standard Brownian motion, $J(\tau) =_d N(0_{k\times 1},\tau (1 - \tau)I_r)$, where $=_d$ signifies equality in distribution. Hence, for a given $b$, $[J(\tau)-\tau J(1)]'[J(\tau)-\tau J(1)]/\tau(1-\tau) =_d \chi^2(r)$, which in turn implies that
\begin{align}
W(b) \to_d \chi^2(r),
\end{align}
as $N,\,T\to\infty$ with $T/N\to 0$. This result holds for all $b$ satisfying $\tau \in \mathcal{T} \subset (0,1)$, including $b_0$.

\begin{remark}
It is important to note that while we do require $T$ to be large, $N$ is still the most important dimension of the data. In the online appendix, we use Monte Carlo simulations as a means to evaluate the importance of the large-$T$ requirement. According to the results, the Wald tests perform well even if $T$ is as small as 10. The large-$T$ requirement is therefore not very important in applied work.
\end{remark}

So far we have taken the date of the break as given. If the date of the break is unknown, as it usually is in practice, then $H_0$ can be tested against the alternative hypothesis of a single structural break at some unknown date $b\in \mathcal{B}'$, which we can formulate as $H_{1}: \bigcup_{b\in\mathcal{B}'} H_1(b)$. Many researchers follow Andrews (1993) and take the supremum of Wald test statistics over all possible breakpoints, and therefore so shall we. The test statistic that we will be considering is therefore given by
\begin{equation}
SW = \sup_{b \in \mathcal{B}'} W(b).
\end{equation}
The asymptotic distribution of this test statistic depends on the distribution of $W(b)$, and is presented in the following corollary to Theorem \ref{thm:dist2}.

\begin{corollary}\label{cor:dist2}
Suppose that $H_0$ holds, and that the conditions of Theorem \ref{thm:dist2} are met. Then, as $N,\,T\to \infty$ with $T/N\to 0$,
\begin{equation}
SW \to_d \sup_{\tau \in \mathcal{T}}\frac{[J(\tau)-\tau J(1)]'[J(\tau)-\tau J(1)]}{\tau(1-\tau)}.
\end{equation}
\end{corollary}

The limiting distribution in Corollary \ref{cor:dist2} is the supremum of the square of a standardized tied-down Bessel process of order $r$, which has appeared previously in Andrews (1993), and Hidalgo and Schafgans (2017), among others. The critical values only depend on $r$ and $\epsilon$, and can be found in Table I of Andrews (1993).

The above results rely on the availability of a consistent estimator $\hat \Sigma_\delta(b)$ of the asymptotic covariance matrix of $\hat \delta(b)$, which is given by $\Sigma_\delta = \Omega_V^{-1}\Psi_V\Omega_V^{-1}$. A natural approach in the current large-$T$ setting is to take
\begin{equation}\label{eq:wvar}
\hat \Sigma_\delta(b)= \hat \Omega_V(b)^{-1}\hat \Psi_V(b)\hat \Omega_V(b)^{-1},
\end{equation}
where
\begin{align}
\hat \Omega_V(b) & = (NT)^{-1}\tilde{Z}(b)'M_{\tilde{X}}\tilde{Z}(b),\\
\hat \Psi_V(b) & = \hat \Psi_{V,0}(b) + \sum_{j=1}^{T-1}k\left(\frac{j}{S_T}\right)(\hat \Psi_{V,j}(b) + \hat \Psi_{V,j}(b)'),\\
\hat \Psi_{V,j}(b) & =\frac{1}{NT}\sum_{i=1}^{N} \sum_{t=j+1}^{T}\hat \varepsilon_{i,t}(b)\hat \varepsilon_{i,t-j}(b) \tilde z_{i,t}(b)\tilde z_{i,t-j}(b)'.
\end{align}
Here $k(\cdot)$ is a real-valued kernel, $S_T$ is the bandwidth parameter, $\hat \varepsilon_{i,t}(b)$ is the $t$-th row of the $T\times 1$ vector $\hat \varepsilon_i(b) = (\hat \varepsilon_{i,1}(b),...,\hat \varepsilon_{i,T}(b))'$, which is in turn the $i$-th block of the $NT \times 1$ vector $\hat \varepsilon(b) = (\hat \varepsilon_1(b)',...,\hat \varepsilon_N(b)')' = M_{\tilde{X}(b)} (\tilde{Y}(b)-\tilde{Z}(b)\hat{\delta}(b))$, and the $1\times r$ vector $\tilde z_{i,t}(b)'$ is the corresponding row of the $NT\times r$ matrix $M_{\tilde{X}(b)}\tilde{Z}(b)$. Alternatively, $\hat \Sigma_\delta(b)$ may be estimated non-parametrically, as in Pesaran and Tosetti (2011).

\begin{remark}
In the special case when $\varepsilon_{i,t}$ is independently and identically distributed across both $i$ and $t$ with variance $\sigma_\varepsilon^2$, $\Sigma_\delta$ reduces to $\Sigma_\delta = \sigma_\varepsilon^2 \Omega_V^{-1}$, which can in turn be estimated using
\begin{align}
\hat \Sigma_\delta(b) = \hat \sigma_\varepsilon^2(b)\hat \Omega_V(b)^{-1},
\end{align}
where $\hat \sigma_\varepsilon^2(b) = (NT)^{-1}\hat \varepsilon(b)'\hat \varepsilon(b)$.
\end{remark}

Once the presence of a break has been established and its location determined, it is possible to make inference regarding $\theta = (\beta',\delta')'$. Let us therefore define $w_{i,t}(b) = (x_{i,t}',z_{i,t}(b)')'$, such that \eqref{eq:y} can be written as
\begin{equation}
y_{i,t}= \theta'w_{i,t}(b_0)+e_{i,t}.
\end{equation}
The CCE estimator of $\theta$ is given by $\hat{\theta} = \hat{\theta}(\hat b)$, where
\begin{align}
\hat{\theta}(b) = (\tilde{W}(b)'\tilde{W}(b))^{-1} \tilde{W}(b)'\tilde{Y},
\end{align}
with $\tilde W(b) = (\tilde W_1(b)',...,\tilde W_N(b)')'$ being $NT\times (k+r)$, $\tilde{W}_i(b) = M_{\bar{H}(b)} W_i$ as before, and
$W_i(b) = (w_{i,1}(b)',...,w_{i,T}(b)')'$ being $T\times (k+r)$. By using the results of Westerlund et al. (2019), we can show that under the conditions of Section \ref{sect:est} with $\tilde W(b)$ in place of $\tilde X$, as $N\to\infty$,
\begin{align}
\sqrt{N}(\hat{\theta} -\theta)|F \to_d N(0_{(k+r)\times 1}, \Omega_W^{-1}\Psi_W\Omega_W^{-1}),
\end{align}
where
\begin{align}
\Omega_W &= \lim_{N\to\infty}\frac{1}{N}\sum_{i=1}^N E(\tilde W_{i}(b_0)'\tilde W_{i}(b_0)|F), \\
\Phi_W &= \lim_{N\to\infty}\frac{1}{N}\sum_{i=1}^N E(\tilde W_{i}(b_0)'\Sigma_{\varepsilon,i}\tilde W_{i}(b_0)|F).
\end{align}
Hence, $\sqrt{N}(\hat{\theta} -\theta)$ is asymptotically normal conditionally on $F$, which means that it supports standard normal and chi-squared inference. If $ T\to \infty $, then the asymptotic distribution of $ \hat\theta $ is the one given by Theorem 4 of Pesaran (2006).

\section{Application stock market reaction of COVID--19}\label{sect:application}

\subsection{Motivation}

COVID--19 broke out in China in December 2019. Roughly one year later, WHO (2021) reports 94 million confirmed cases and over two million deaths. We also know that because of lockdowns, travel restrictions and social distancing policies, in 2020 GDP dropped by 4.2\% globally and real world trade contracted by 10.3\% (OECD, 2020).\footnote{By comparison, the lowest global GDP growth rate during the 2007--2009 global financial crisis was $-1.7$\% in 2009.} The economic impact of the pandemic has therefore been substantial. This is what we know. There are some signs of recovery in the years to come; however, the global outlook is extremely uncertain, even in the short term. As an indication of this, the OECD world GDP projections for 2021 ranges from $-2.75$\% to $5$\%, depending on, among other things, the evolution of the pandemic, the actions taken to contain the spread of the virus and their economic impact, and the time until effective vaccines can be deployed. Hence, even now, one year after the outbreak, much is uncertain.

The uncertainty we face today is nothing compared to one year ago. At this time, little was known about the new virus, but it was clear that it was very infectious and deadly, as, in contrast to previous infectious disease outbreaks, most countries begun to announce the number of cases and deaths on a daily basis. Many were chocked by how quickly these numbers were increasing. Governments scrambled with emergency actions, such as closing schools and workplaces, travel bans, or even complete curfews, to try to contain the spread. However, since their effectiveness was far from clear and they made it impossible for firms and workers to continue their operations without knowing if and when they would be compensated, these actions added to the already existing uncertainty, leading to widespread public fear (see Mamaysky, 2020, and Phan and Narayan, 2020). This was visibly apparent with news coming in of supermarkets being stocked out of toilet paper (Aggarwal et al., 2020).

In times of extreme uncertainty, stock markets often respond dramatically to news about the underlying economic and market conditions (see Mamaysky, 2020). This is what happened during the global financial crisis of 2007--2009 and it happened again in the initial stages of the pandemic. In January 2020, the news reporting was comparable to what it was in the beginning of the SARS (severe acute respiratory syndrome) and Ebola epidemics. By February, however, COVID--19 started to dominate newspaper discussions of the economy, and by March, almost all such discussions were about COVID--19 (Baker et al., 2020).\footnote{Not all news were about the economy and many were just rumors, but they still attracted considerable attention and were therefore important in setting the public sentiment at the time. For example, on February 17, a run on toilet paper in Hong Kong was mentioned for the first time, and became a highly contagious story. Some people in locked-down China reportedly were reduced to searching for minnows and ragworms to eat. In Italy, there were stories of medical workers in overwhelmed hospitals being forced to choose which patients would receive treatment (Shiller, 2020).} Stock markets responded violently. On March 16, the Chicago Board Option Exchange's volatility index, the so-called ``VIX'', surged past the prior all-time peak reached during the global financial crisis more than a decade ago. The second-worst day ever of the Dow Jones industrial index happened on March 16, and three of the 15 worst days ever of the US market occurred between March 9 and 16 (Wagner, 2020). Stock markets all over the world reacted similarly.

The unprecedented stock market behaviour in the initial stage of COVID--19 has attracted considerable attention not only in the news but also in research. The bulk of the evidence seem to suggest that stock markets have generally responded negatively, although the channel through which this effect works is still largely unknown. Ashraf (2020a) uses data for 64 countries and finds that stock prices have reacted negatively to the pandemic, but only when measured by the number of confirmed cases, as opposed to the death count. This is largely in agreement with the results of Erdem (2020). Ashraf (2020b) employs data for 77 countries. He finds that the COVID-19 effect operates not only through the number of cases, but also through government actions, such as social distancing measures, containment and health responses, and economic support packages. Similar findings have been reported by Aggarwal et al. (2020), and Capelle-Blancard and Desroziers (2020).\footnote{Many studies focus on single countries. There are also those that focus on the volatility of stock returns, as opposed to stock returns themselves. These are not reviewed here.}

The purpose of the current application is to contribute to the above mentioned literature. This is done in three ways. First, we account for the rebound of returns. Faced with near economic collapse, starting with the Federal Reserve's decision on March 16 to buy USD 700 billion worth of US Treasury bonds and mortgage-backed securities, central banks around the world announced aggressive quantitative easing programs (see Hartley and Rebucci, 2020). These announcements were followed by an abrupt increase in stock prices. The US S\&P500 stock market index, for example, increased by 29\% between March 24 and April 17, a surge that left the index back where it stood in August of 2019 when the US economy was booming. The fact that this rebound took place while the pandemic was still ongoing is suggestive of a structural break. Most studies ignore this. The only exceptions known to us are Capelle-Blancard and Desroziers (2020), Mamaysky (2020), and Ramelli and Wagner (2020), who divide their samples into sub-periods based on major events. The breaks are therefore treated as known, which is risky, as misplaced breaks are just as problematic as omitted breaks. The second contribution is that we account for general forms of unobserved heterogeneity. Many studies recognize the problem but assume that it can be handled using country and period fixed effects. However, fixed effects do not work in general when pair-wise cross-section covariances of the regression errors differ across countries, and there is plenty of evidence to support this (see, for example, Zhang et al., 2020). Capelle-Blancard and Desroziers (2020) use robust standard errors but they can only handle weak cross-section dependence. The third contribution is that we account for the smallness of $T$. As alluded to in the previous paragraph, with COVID--19 being such a recent phenomenon, studies of it are constrained to data sets with short time span (see, for example, Salisu and Vo, 2020, for a discussion). Some ``compensate'' by using a relatively high frequency, such as daily data, but not all. To take an extreme example, Aggarwal et al. (2020) use monthly data from December 2019 to May 2020, which means that $T=6$. Moreover, even if data are daily, the sub-periods considered are very short. It is therefore important to use appropriate small-$T$ techniques.

\subsection{Data}

Our dependent variable is stock returns (RET), which we compute as the log difference of the price index.\footnote{We experimented using excess returns. However, because the results were qualitatively the same, and since the previous literature focuses almost exclusively on raw returns, here we only report the results based on using raw returns as the dependent variable.} We utilize four control variables; the US Dollar exchange rate (ER), stock market volatility (VOL), which we proxy using the Chicago Board Options Exchange's CBOE volatility index, world market returns (MRET), as measured by the cross-country average of RET, and the US three-month Treasury bill rate (TBILL) (see, for example, Aggarwal et al., 2020, Capelle-Blancard and Desroziers, 2020, Mamaysky, 2020, and Salisu and Vo, 2020, for similar control variable lineups). ER is motivated by the theoretical work of Dornbusch and Fischer (1980), which says that exchange rates will influence stock returns because they capture the value of firms' future cash flows. VOL can be motivated in part by its ability to predict returns (see, for example, Bollerslev et al., 2009, and Bollerslev et al., 2015), in part by the theory of Glasserman et al. (2020), according to which information shocks can lead to large drops in stock prices and increases in volatility. TBILL captures both the risk-free interest rate and the importance of the US in shaping stock markets around the world. The need to control for MRET is due to CAPM.

We use all available measures of COVID--19 that have sufficient time series data. A total of six variables meet this criterion. The first two capture the spread of the virus. They are the number of confirmed cases (CASE) and deaths (DEATH). The next four variables are indices that capture government response to COVID--19; a government stringency index (STR), a containment and health index (CONT), a government economic support index (ECON), and an overall government response index (RESP). STR records the strictness of government policies that primarily restrict people's behaviour, such as school and workplace closures, stay-at-home requirements, and travel bans. CONT captures mainly social distancing restrictions, but also health system policies such as testing policy, contact tracing, short term investment in healthcare and investments in vaccine. ECON is an index that captures government income support and debt relief. RESP captures all of the above. All indices are on a scale of zero to 100 with a larger value indicating greater stringency, greater commitment to health, greater economic support, and greater overall government response. All data are obtained from Datastream, except for TBILL, which is from Federal Reserve Bank St Louis. The data are weekly and cover $N=61$ countries.\footnote{The included countries are Argentina, Australia, Austria, Belgium, Brazil, Bulgaria, Canada, Chile, China, Croatia, Cyprus, Czech Republic, Denmark, Egypt, Estonia, Finland, France, Germany, Greece, Hong Kong, Hungary, Iceland, India, Indonesia, Ireland, Israel, Italy, Jamaica, Japan, Jordan, Kenya, Kuwait, Latvia, Luxembourg, Malaysia, Mexico, Morocco, the Netherlands, New Zealand, Norway, Oman, Pakistan, Peru, the Philippines, Poland, Portugal, Romania, Russia, Singapore, Slovakia, Slovenia, South Africa, South Korea, Spain, Sri Lanka, Sweden, Switzerland, Thailand, Tunisia, Turkey, and the United Kingdom.} As in many other empirical scenarios, the number of time periods is limited and cannot be increased other than by the passage of time. We take the largest sample period available to us, which covers $T= 38$ weeks, from January 3 to September 25, 2020. The smallness of $T$ in this case means that it is important to use techniques that work even if $T$ is not large. The Monte Carlo results reported in the online appendix suggest that the proposed toolbox should work well here.

\subsection{Implementation}

Both theory and empirical observations stress the importance of news (see, for example, Mamaysky, 2020). We therefore follow the bulk of the existing literature and express all regressors in innovation form by taking first differences. Ashraf (2020b) considers the same stringency, containment and health, and economic support indices as we do. While he includes all three indices at the same time, we do not. The reason is that STR, CONT and RESP are highly collinear with correlations that range from 0.949 to 0.978 (see Capelle-Blancard and Desroziers, 2020, for a similar argument). We therefore include them one at a time.

VOL, MRET and TBILL do not vary by country but only by week. We therefore want to treat these are observed common factors, a possibility that we did not consider in Sections \ref{sect:mod}--\ref{sect:tests}. As pointed out in Section \ref{sect:mod}, the type of factors that can be permitted under Assumption \ref{ass:f} is very broad. This suggests that there is no need to discriminate between known and unknown factors, but that one can just as well treat them all as unknown to be estimated from the data. In fact, this is the main rationale for writing \eqref{eq:y} and \eqref{eq:x} in terms of (the unknown) $f_t$ only. The main drawback of this fully unknown factor treatment is that it puts strain on the Assumption \ref{ass:load} condition that $m\leq k$, as $k$ is fixed and additional factors increase $m$ even if they are known. For this reason, it may be preferable to be able to distinguish between known and unknown factors. Fortunately, in CCE this is very easy. Let us therefore assume that there are two sets of factors, $f_t$ and $d_t$, where $f_t$ is a $m\times 1$ vector of unknown factors, just as before, while $d_t$ is a $n\times 1$ vectors of known common regressors, which in this section is comprised of a constant (country fixed effects), VOL, MRET and TBILL. Hence, now the total number of factors is equal to $m+n$, out of which $n$ are known. The model is the same as before but with \eqref{eq:y} and \eqref{eq:x} replaced by
\begin{align}
y_{i,t} & = \alpha'_i d_t + \beta'x_{i,t}+\delta'z_{i,t}(b)+e_{i,t}, \label{eq:yd}\\
x_{i,t} & = A_i'd_t+\Gamma_i'f_t+v_{i,t},
\end{align}
where $\alpha_i$ and $A_i$ are $n\times 1$ and $n\times k$, respectively. Provided that Assumption \ref{ass:load} holds, so that $\mathrm{rank}(\bar \Gamma)=m \leq k$, similarly to \eqref{eq:fsol}, we can show that $(d_t',\bar x_t')'$ is rotationally consistent for $f_t$. Hence, instead of using $M_{\bar X}$ to purge the effect of $F$ in the estimation of $b$, $M_{(D,\bar X)}$ must be used, where $D=(d_1,...,d_T)'$ is $T\times n$. This is, however, the only change needed in order to accommodate the known regressors in $d_t$. Note in particular how the condition that $m \leq k$ is unaffected by $n$. Similarly, the only change needed in order to account for the known factors in the break tests is to replace $M_{\bar H(b)}$ with $M_{(D,D(b)\bar H(b))}$, where $D(b)=(d_1\mathbb{I}(1>b),...,d_T\mathbb{I}(T>b))'$.

The above discussion suggests that in terms of the known factor-augmented version of \eqref{eq:y} in \eqref{eq:yd}, in this section $y_{i,t}$ is RET, $d_t$ is a constant, VOL, MRET and TBILL, and $x_{i,t}$ is ER, CASE, DEATH, ECON, and one of STR, CONT and RESP. We allow the coefficients of the COVID--19 spread and response variables to be breaking, but not the coefficient of ER. The date of the break is treated as unknown not only in the estimation but also on the testing. We therefore focus on the $SW$ test, which we implement using $15$\% trimming ($\epsilon=0.15$) (as in, for example, Andrews, 1993, and Bai, 1997a). Based on its good performance in the Monte Carlo study reported in the online appendix, the asymptotic covariance matrix of $\hat \delta(b)$ is computed based on the Bartlett kernel with the bandwidth parameter $S_T$ set equal to $S_T=\lfloor T^{1/3} \rfloor$.

As in the present paper, Bai (2010) focuses on the single break case, although he also discuss the possibility of having multiple breaks. As he remarks, if the number of breaks is given, the one-at-a-time approach of Bai (1997b) can be used to estimate the breakpoints, and if the number of breaks is unknown, a test for existence of a break can be applied to each subsample before estimating another breakpoint. The same approach can be used also in the current more general context. The results are discussed in the next section.

\subsection{Results}

Following the convention in the literature (see, for example, Ashraf, 2020a, 2020b, Capelle-Blancard and Desroziers, 2020, and Erdem, 2020), Table 1 reports the mean, standard deviation, minimum and maximum of each variable. RET has a mean value of $-0.392$ with a standard deviation of $4.44$. The fact that the mean is negative indicates that the pandemic has affected stock markets negatively. The mean of CASE and DEATH are positive, as expected because the pandemic has not settled down yet. The results for the response variables show that governments have responded to the pandemic. In order to get a feeling for the validity of the conventional fixed effects assumption, we computed the CD test of Pesaran (2021), which tests the null hypothesis of no remaining cross-sectional correlation after controlling for fixed effects. The null is rejected at all conventional significance levels for all variables, suggesting that, as expected given the above discussion, fixed effects are not enough to account for the cross-section correlation. The results of the unit root tests of Elliott et al. (1996), and Pesaran (2007) confirm that all the variables are stationary.\footnote{The results of the CD test motivate the use of the CIPS test of Pesaran (2007), which allows for cross-section correlation. The use of the test of Elliott et al. (1996) is motivated by its good power properties.}

\begin{table}[]
\caption{Descriptive statistics.}
\hskip 12 pt
\centering
\footnotesize
\begin{tabular}{lllllll}
\hline\hline
Variable           & Mean    & SD & Min     & Max   & UR    & CD  \\
\hline
RET              & $-0.392$  & $4.440$     & $-26.795$ & $17.656$ & $-5.769$*** & $161.010$*** \\
CASE             & $99.125$  & $1314.487$  & $-20289$  & $28789$  & $-4.137$*** & $13.631$***  \\
DEATH            & $275.842$ & $985.860$   & $-1507$   & $8101$   & $-2.098$*   & $63.246$***  \\
STR              & $1.348$   & $9.163$     & $-47.2$   & $88.9$   & $-5.768$*** & $114.333$*** \\
RESP             & $1.543$   & $6.732$     & $-30.3$   & $64.3$   & $-5.633$*** & $117.615$*** \\
CONT             & $1.501$   & $7.207$     & $-35.4$   & $70.8$   & $-5.651$*** & $108.350$*** \\
ECON             & $1.796$   & $10.293$    & $-50 $    & $100$   & $-5.263$*** & $55.289$***  \\
ER               & $0.460$   & $41.260$    & $-740$    & $1155$   & $-5.930$*** & $4.418$***   \\
VOL              & $1.663$   & $20.841$    & $-33.678$ & $85.372$ & $-4.160$*** &   $-$      \\
TBILL            & $0.399$   & $0.535$     & $0.100$   & $1.520$  & $-3.405$**  &    $-$     \\
MRET             & $-0.392$  & $3.515$     & $-14.864$ & $5.479$  & $-4.312$*** &    $-$     \\
\hline\hline
\end{tabular}
\begin{tablenotes}
\item \emph{Notes}: ``Mean'', ``SD'', ``Min'' and ``Max'' refer to the sample average, the standard deviation, the minimum value and the maximum value of each variable. The column labelled ``UR'' reports some unit root test results. If the variable exhibits cross-sectional variation we employ the CIPS test of Pesaran (2007), which allows cross-sectional dependence in the form of a common factor. If the variable only varies over time, we employ unit root test of Elliott et al. (1996). Both tests are augmented with four lags to capture any serial correlation in the regression errors. The column labelled ``CD'' reports the results obtained by applying Pesaran's (2004) test for cross-sectional correlation. The CD results for VOL, TBILL and MRET are not reported as these variables do not vary by country. The variables are stock returns (RET), cases (CASE), deaths (DEATH), government stringency (STR), overall government response (RESP), government containment and health (CONT), government economic support  (ECON), the US dollar exchange rate (ER), stock market volatility (VOL), the three-month US Treasury Bill rate (TBILL) and average stock returns (MRET).																										\end{tablenotes}
\end{table}

Table 2 contains our main results. It reports the estimated coefficients and their significance, the $SW$ test values and their significance, the estimated break date, and the associated 95\% break date confidence interval. The table only contains the results for the COVID--19 variables, which are our main regressors. The controls are included but we do not report their results. Three specifications are considered, one for each of STR, CONT and RESP.

The first thing to note is that the $SW$ test is highly significant in all three specifications and that the break date is estimated to the first week of April, which starts on March 30 and ends on April 5. This is consistent with the quantitative easing interventions of major central banks and the sharp stock market rise that followed. For example, the S\&P 500 stock market index lost 34\% of its value between February 19 and March 23, but abruptly regained the vast majority of this loss, rising 29\% between March 24 and April 17. Stock markets all around the world experienced similar surges (see International Monetary Fund, IMF, 2020). There may be confounding factors that may have affected stock returns positively. However, we note that the quantitative easing announcements were among the largest news items at the time (Mamaysky, 2020). A full table of announcements can be found in Hartley and Rebucci (2020). The dates of some notable announcements in March 2020 are the European Central Bank on the 18th, the Bank of England and the Reserve Bank of Australia on the 19th, the Reserve Bank of New Zealand on the 23rd, the Bank of Korea on the 25th, the Federal Reserve and the Bank of Canada on the 27th. Most of the emerging economies' central banks made their announcements in the last 10 days of March. Our estimated break date is located directly after these announcements. Quantitative easing pushes interest rates down and this has two possible effects, which both result in an increase in stock prices (see, for example, Bernanke, 2012). First, by decreasing the discount rate, quantitative easing increases the present value of future cash flows. Second, quantitative easing makes relatively safe assets unattractive, which creates an incentive for investors to rebalance their portfolios to include more stocks, and this in turn pushes stock prices up. We therefore speculate that it was the quantitative easing announcements that caused the break in the stock return-COVID--19 relationship.\footnote{We also note that our estimated breakpoint does not coincide with the sample splits considered by Capelle-Blancard and Desroziers (2020), Mamaysky (2020), and Ramelli and Wagner (2020).} As explained earlier in this section, the $SW$ test was applied not only to the full sample but also to the pre- and post-break periods. There were, however, no significant breaks in the pre- and post-break periods, and so we conclude that there is just one break.

\begin{table}[]
	\caption{Main results.}
	\hskip 12 pt
	\centering
	\footnotesize
	\begin{tabular}{lclll}
		\hline\hline
		Regressor & Coeff                 & Spec 1          & Spec 2     & Spec 3   \\ \hline
		CASE & $\beta$                               & $-0.001$**                                  & $-0.002$***        & $-0.002$***      \\
		& & (0.000)                                 & (0.001)    & (0.001)    \\
		& $\delta$                              & 0.001***                                     & 0.002***         & 0.002***         \\
		& & (0.000)                                & (0.001)    & (0.001)    \\
		DEATH & $\beta$                              & $-0.002$*                                    & $-0.002$**        & $-0.002$**        \\
		& & (0.001)                                  & (0.001)     & (0.001)     \\
		& $\delta$                             & 0.002*                                     & 0.002**         & 0.002**         \\
		& & ( 0.001)                                 & (0.001)     & (0.001)     \\
		ECON & $\beta$                               & 0.056**                                     & 0.070***         & 0.060**         \\
		& & (0.028)                                 & (0.027)    & (0.026)     \\
		& $\delta$                              & $-0.068$**                                    & $-0.082$***        & $-0.073$***        \\
		& & (0.029)                                 & (0.027)    & (0.027)    \\
		STR & $\beta$                                & $-0.077$***                                    &               &               \\
		& & ( 0.020)                               &               &               \\
		& $\delta$                               & 0.075***                                     &               &               \\
		& & ( 0.022)                               &               &               \\
		GOV & $\beta$                                &                                           & $-0.075$***        &               \\
		& &                                           & (0.028)    &               \\
		& $\delta$                               &                                           & 0.066**         &               \\
		& &                                           & (0.031)     &               \\
		CONT & $\beta$                               &                                           &               & $-0.065$***        \\
		& &                                           &               & (0.023)    \\
		& $\delta$                              &                                           &               & 0.056**         \\
		& &                                           &               & (0.024)     \\ \hline
		$r$                             &                & 4                                         & 4             & 4             \\
		$n$                             &                 & 4                                         & 4             & 4             \\
		$SW$                             &                & 31.182***                                 & 24.108***     & 24.096***     \\
		$\hat b$                         &              & Apr 5                                    & Apr 5        & Apr 5       \\
		95\% CI                           &               & [Mar 29,Apr 12]                             & [Mar 29,Apr 12] & [Mar 29,Apr 12] \\ 		\multicolumn{5}{l}{Unreported controls: Country fixed effects, VOL, TBILL, MRET and ER}             \\ 
\hline\hline                                       	
\end{tabular}
\begin{tablenotes}																					
\item  \emph{Notes}: The dependent variable is stock returns (RET). The main regressors are cases (CASE), deaths (DEATH), government stringency (STR), overall government response (RESP), government containment and health (CONT) and government economic support (ECON). All specifications include country fixed effects, the US dollar exchange rate (ER), stock market volatility (VOL), the three-month US Treasury Bill rate (TBILL) and average stock returns (MRET) as controls. As estimates of the unobserved factors we use the main regressors of each specification and ER. All specifications also include country fixed effects, VOL, TBILL and MRET as observed common factors. $\beta$ and $\delta$ refer to the pre-break slope and the size of the break, respectively. $r$ and $n$ refer to the number of panel data regressors and observed common factors, respectively. $SW$ refers to the sup-Wald test for the existence of a structural break, $\hat b$ refers to the estimated breakpoint and ``95\% CI'' refers to the associated $95$\% confidence interval. The reported dates refer to the last day of the relevant week. The numbers within parentheses are the standard errors. Finally, *, ** and *** denote statistical significance at the $10$\%, $5$\% and $1$\% levels, respectively.
\end{tablenotes}
\end{table}

Another observation is that all the COVID--19 regressors enter significantly but only before the break. Specifically, the estimated pre-break coefficients ($\beta$) are all significant, as are the estimated breaks ($\delta$), but they sum up to zero, and the sum is insignificant in all cases. In other words, the estimated post-break effects ($\beta+\delta$) are insignificant. Consider CASE and DEATH. Their pre-break effect is significantly negative, which is consistent with existing results (see, for example, Ashraf, 2020a, 2020b, Capelle-Blancard and Desroziers, 2020, and Erdem, 2020). Hence, as expected, stock markets therefore initially responded negatively to the news of the outbreak of the virus. This negative effect is, however, completely eliminated by the break, which is estimated to be of the same magnitude but of opposite sign. The post-break effect of CASE and DEATH is therefore estimated to zero, suggesting that the central bank interventions have had a substantial positive effect on stock markets.

Let us now move on to the response regressors, STR, CONT, ECON and RESP. The estimated pre-break effect of ECON is significantly positive, meaning that stock markets initially responded positively to news of increased government support, which is again in accordance with our a priori expectations. After the break, however, stock markets became insensitive to such news. Similarly, while initially markets responded negatively to announcements of stricter and more extensive government restrictions, as measured by STR and RESP, after the break they did not respond at all. The same is true for CONT, which is probably due to the fact that while this variable captures both social distancing restrictions and investments in healthcare, the restrictions are weighted higher in the construction of the index and they did came first. As an illustration of the effect of STR, CONT and RESP, in Figure 1 we plot the cross-sectional averages of these variables against that of RET. We see that while before the break the co-movement between average RET on the one hand and STR, CONT and RESP on the other hand is clearly negative, after the break the co-movement is much weaker. These results are quite different from existing ones. Capelle-Blancard and Desroziers (2020) find that STR has a positive but insignificant effect, which becomes significant only in the absence of fixed effects or other control variables. Ashraf (2020b) reports a significantly negative effect of STR, a significantly positive effect of CONT and an insignificant effect of ECON. However, these other studies only allow for fixed effects and they do not take into consideration our estimated breakpoint, which could very well explain the observed differences in the results.

\begin{figure}[]
	\caption{Plotting the cross-sectional averages of RET, STR, CONT and RESP.}
	\centering
	\includegraphics[width=14cm,keepaspectratio]{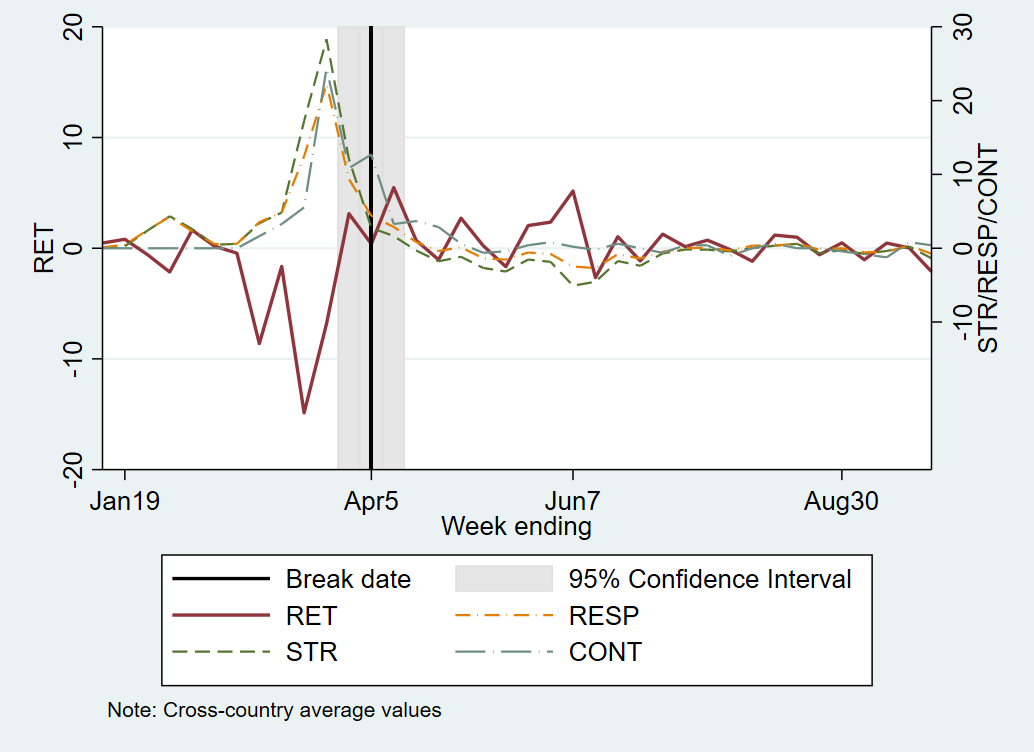}
	\caption*{\footnotesize \emph{Notes:} The figure plots the weekly cross-sectional averages of stock returns (RET), government stringency (STR), government containment and health (CONT) and overall government response (RESP). The break date estimate and the associated $95$\% confidence interval are taken from Table 2. The reported dates refer to the last day of the relevant week.}
\end{figure}

According to the results of Baker et al. (2020), and Mamaysky (2020), in the early phase of the pandemic (late February to late March) stock market movements were driven by news about the virus. In fact, markets were ``hypersensitive'' and overreacted not only to news themselves but also to other markets' reaction to news (Mamaysky, 2020). ``Markets started to oscillate wildly, and people suddenly realized that the virus could affect them directly. Panic selling in the stock market went hand-in-hand with panic buying in supermarkets'' (Wagner, 2020, page 440). This explains why initially stock markets reacted significantly to all COVID--19 related news (CASE, DEATH, STR, CONT, ECON and RESP). The powerful central bank interventions acted as a wake-up call. They signalled a clear commitment to deal with the pandemic, thereby bringing some certainty to an otherwise extremely uncertain future. Stock markets reacted positively and progressed on a path to recovery. This is noteworthy because the economic conditions have been steadily deteriorating as a result of closures and social distancing (see IMF, 2020). As Krugman (2020) puts it, ``[t]he relationship between stock performance -- largely driven by the oscillation between greed and fear -- and real economic growth has always been somewhere between loose and nonexistent''.

The explanation for our results given in the previous paragraph is consistent with (at least) two theories. The first is the so-called ``overreaction'' hypothesis of Daniel et al. (1998), and Hong and Stein (1999), which states that investors overreact to negative shocks, such as those that hit stock markets in the early phase of the pandemic. As more information becomes available, however, and the central bank announcement were very informative, investors correct their behavior, which leads to market recovery. The second theory is that of Glasserman et al. (2020). It states that information shocks, such as the outbreak of COVID--19, can lead to large drops in prices and increases in volatility, which in turn cause prices to become hypersensitive to newsflow. However, information can also push prices out of hypersensitivity, and our results show that in the post-break regime returns are no longer reacting to news of the pandemic.

\section{Conclusions}\label{sect:concl}

The main aim of this paper is to provide a toolbox that meets the basic needs of researchers interested in a linear panel data model with a possible structural break. The toolbox allows researchers to test for the presence of a break, and, if a break is detected, to also estimate the location of the break and construct a confidence interval for the true breakpoint. The toolbox does not require that the data are independent, nor that $T$ is large, which means that it widely is applicable. 

The new toolbox in employed to investigate the relationship between stock market returns and COVID--19 in a sample covering 61 countries across 38 weeks. Stock markets all over the world plunged in the early phase of the pandemic but they quickly rebounded, and this rebound took place although the end of the pandemic is still not in sight. Our analysis shows that while initially responsive, the effect of COVID-19 stopped dead at the end of March-beginning of April 2020. We attribute this break to the massive quantitative easing programs announced by central banks around the world in the second half of March.

\newpage

\section*{References}

\begin{description}
\item Aggarwal, S., S. Nawn and A. Dugar (2020). What Caused Global Stock Market Meltdown during the COVID Pandemic--Lockdown Stringency or Investor Panic? \emph{Finance Research Letters} \textbf{31}, 101690.

\item Andrews, D. W. K. (1993). Tests for Parameter Instability and Structural Change with Unknown Change Point. \emph{Econometrica} \textbf{61}, 821--856.

\item Antoch, J., J. Hanousek, L. Horv\'{a}th, M. Hu\v{s}kov\'{a} and S. Wang (2019). Structural Breaks in Panel Data: Large Number of Panels and Short Length Time Series. \emph{Econometric Reviews} \textbf{38}, 828--855.

\item Ashraf, B. N. (2020a). Stock Markets' Reaction to COVID-19: Cases or Fatalities? \emph{Research in International Business and Finance} \textbf{54}, 101249.

\item Ashraf, B. N. (2020b). Economic Impact of Government Interventions during the COVID-19 Pandemic: International Evidence from Financial Markets. \emph{Journal of Behavioral and Experimental Finance} \textbf{27}, 100371.

\item Bai, J. (1997a). Estimation of a Change Point in Multiple Regression Models. \emph{Review of Economics and Statistics} \textbf{79}, 551--563.

\item Bai, J. (1997b). Estimating Multiple Breaks One at a Time. \emph{Econometric Theory} \textbf{13}, 315--352.

\item Bai, J. (2010). Common Breaks in Means and Variances for Panel Data. \emph{Journal of Econometrics} \textbf{157}, 78--92.

\item Baker S. R., N. Bloom, S. J. Davis, K. J. Kost, M. C. Sammon and T. Viratyosin (2020). The Unprecedented Stock Market Impact of COVID-19. \emph{Covid Economics} \textbf{1}, 33--42.

\item Baltagi, B. H., Q. Feng and C. Kao (2016). Estimation of Heterogeneous Panels with Structural Breaks. \emph{Journal of Econometrics} \textbf{191}, 176--195.

\item Baltagi, B. H., C. Kao and L. Liu (2017). Estimation and Identification of Change Points in Panel Models with Nonstationary or Stationary Regressors and Error Term. \emph{Econometric Reviews} \textbf{36}, 85--102.

\item Bernanke, B. (2012). Monetary Policy Since the Onset of the Crisis. Remarks at the Federal Reserve Bank of Kansas City Economic Symposium, Jackson Hole, Wyoming, August 31.

\item Boldea, O., B. Drepper and Z. Gan (2020). Change Point Estimation in Panel Data with Time-Varying Individual Effects. Forthcoming in \emph{Journal of Applied Econometrics}.

\item Bollerslev, T., G. Tauchen and H. Zhou (2009). Expected Stock Returns and Variance Risk Premia. \emph{Review of Financial Studies} \textbf{22}, 4463--4492.

\item Bollerslev, T., L. Xu and H. Zhou (2015). Stock Return and Cash Flow Predictability: The Role of Volatility Risk. \emph{Journal of Econometrics} \textbf{187}, 458--471.

\item Capelle-Blancard, G., and A. Desroziers (2020). The Stock Market is not the Economy? Insights from the Covid-19 crisis. \emph{Covid Economics} \textbf{28}, 29--69.

\item Daniel, K., D. Hirshleifer and A. Subrahmanyam (1998). Investor Psychology and Security Market Under- and Overreactions. \emph{Journal of Finance} \textbf{53}, 1839--1886.

\item Dornbusch, R., and S. Fischer (1980). Exchange Rates and the Current Account. \emph{American Economic Review} \textbf{70}, 960--971.

\item Elliott, G., T. J. Rothenberg, and J. H. Stock (1996). Efficient Tests for an Autoregressive Unit Root. \emph{Econometrica} \textbf{64}, 813--836.

\item Erdem, O., (2020). Freedom and Stock Market Performance during COVID-19 Outbreak. \emph{Finance Research Letters} \textbf{36}, 101671.

\item Glasserman, P., H. Mamaysky, and Y. Shen (2020). Dynamic Information Regimes in Financial Markets. Working paper.

\item Hartley, J., S., and A. Rebucci, (2020). An Event Study of COVID--19 Central Bank Quantitative Easing in Advanced and Emerging Economies. NBER Working Paper 27339.

\item Hidalgo, J., and M. Schafgans (2017). Inference and Testing Breaks in Large Dynamic Panels with Strong Cross Sectional Dependence. \emph{Journal of Econometrics} \textbf{196}, 259--274.

\item Hong, H., Stein, J., (1999). A Unified Theory of Underreaction, Momentum Trading, and Overreaction in Asset Markets. \emph{Journal of Finance} \textbf{54}, 2143--2184.

\item IMF (2020). Special Series on COVID-19: The Disconnect between Financial Markets and the Real Economy. Washington.

\item Karabiyik, H., S. Reese and J. Westerlund (2017). On the Role of the Rank Condition in CCE Estimation of Factor-Augmented Panel Regressions. \emph{Journal of Econometrics} \textbf{197}, 60--64.

\item Kim, D. (2011). Estimating a Common Deterministic Time Trend Break in Large Panels with Cross Sectional Dependence. \emph{Journal of Econometrics} \textbf{164}, 310--330.

\item Kim, D. (2014). Common Local Breaks in Time Trends for Large Panels. \emph{Econometric Journal} \textbf{17}, 301--337.

\item Krugman, P. (2020). Crashing Economy, Rising Stocks: What's going on? New York Times, April 30.

\item Li, D., J. Qian and L. Su (2016). Panel Data Models With Interactive Fixed Effects and Multiple Structural Break. \emph{Journal of the American Statistical Association} \textbf{111}, 1804--1819.

\item OECD (2020). Economic Outlook, Issue 2, OECD Publishing, Paris.

\item Pesaran, M. H. (2006). Estimation and Inference in Large Heterogeneous Panels with a Multifactor Error Structure. \emph{Econometrica} \textbf{74}, 967--1012.

\item Pesaran, M.H. (2007). A Simple Panel Unit Root Test in the Presence of Cross-Section Dependence. \emph{Journal of Applied Econometrics} \textbf{22}, 265--312.

\item Pesaran, H. M. (2021). General Diagnostic Tests for Cross-Sectional Dependence in Panels. Forthcoming in \emph{Empirical Economics}.

\item Pesaran, M. H., and E. Tosetti (2011). Large Panels with Common Factors and Spatial Correlation. \emph{Journal of Econometrics} \textbf{161(2)}, 182--202.

\item Phan, D. H. B., and P. K. Narayan (2020). Country Responses and the Reaction of the Stock Market to COVID--19--a Preliminary Exposition. \emph{Emerging Markets Finance and Trade} \textbf{56}, 2138--2150.

\item Ramelli, S., and A. F. Wagner (2020). Feverish Stock Price Reactions to COVIS-19. Forthcoming in \emph{Review of Corporate Finance Studies}.

\item Salisu, A. A., and  X. V. Vo (2020). Predicting Stock Returns in the Presence of COVID-19 Pandemic: The Role of Health News. \emph{International Review of Financial Analysis} \textbf{71}, 101546.

\item Shiller, R. (2020). Opinion: Robert Shiller Explains the Pandemic Stock Market and why it's Decoupled from the Economy. Project Syndicate, July 11, 2020.

\item Wagner, A. F (2020). What the Stock Market Tells us about the Post-COVID-19 World. \emph{Nature Human Behaviour} \textbf{4}, 440.

\item Westerlund, J. (2019). Common Breaks in Means for Cross--Correlated Fixed-$T$ Panel Data. \emph{Journal of Time Series Analysis} \textbf{40}, 248--255.

\item Westerlund, J., Karabiyik, H., and Narayan, P., (2017). Testing for Predictability in Panels with General Predictors. \emph{Journal of Applied Econometrics} \textbf{32}, 554--574.

\item Westerlund, J., Y. Petrova and M. Norkute (2019). CCE in Fixed-$T$ Panels. \emph{Journal of Applied Econometrics} \textbf{34}, 746--761.

\item WHO (2021). Weekly operational update on COVID--19 - 19 January 2021. Retrieved from: https://www.who.int/publications/m/item/weekly-operational-update-on-covid-19---19-january-2021.

\item Zhang, D., M. Hu, Q. Ji (2020) Financial Markets under the Global Pandemic of COVID-19. \emph{Finance Research Letters} \textbf{36}, 101528.
\end{description}

\end{document}